\documentclass[%
reprint,
showpacs,
bibnotes,
amsmath,amssymb,
aps,
prb,
floatfix,
longbibliography
]{revtex4-1}

\newlength{\myfigwidth}
\usepackage[breaklinks=true,colorlinks,citecolor=blue,linkcolor=blue,urlcolor=blue]{hyperref}

\renewcommand{\BibitemShut}[1]{}

\usepackage{epsfig}
\usepackage{mathrsfs}
\usepackage{amssymb}
\usepackage{color}
\usepackage{csquotes}
\usepackage{latexsym}
\usepackage{subfigure}
\usepackage{float}
\usepackage{marginnote}
\usepackage{dsfont}
\usepackage{tikz}

\newcommand{\bcen}{\begin{center}}
\newcommand{\ecen}{\end{center}}
\newcommand{\btab}{\begin{tabular}}
\newcommand{\etab}{\end{tabular}}
\newcommand{\bdes}{\begin{description}}
\newcommand{\edes}{\end{description}}

\newcommand{\beq}{\begin{equation}}
\newcommand{\eeq}{\end{equation}}
\newcommand{\bea}{\begin{eqnarray}}
\newcommand{\eea}{\end{eqnarray}}

\newcommand{\bary}{\begin{array}}
\newcommand{\eary}{\end{array}}
\newcommand{\benum}{\begin{enumerate}}
\newcommand{\eenum}{\end{enumerate}}
\newcommand{\bitem}{\begin{itemize}}
\newcommand{\eitem}{\end{itemize}}

\newcommand{\mylabel}[1]{\label{#1}} 
\begin{document}
\title{First Principles Prediction of Amorphous Phases Using Evolutionary Algorithms }
\author{Suhas Nahas}\email[]{shsnhs@iitk.ac.in}
\author{Anshu Gaur}\email[]{agaur@iitk.ac.in}
\author{Somnath Bhowmick}\email[]{bsomnath@iitk.ac.in}
\affiliation{Department of Material Science and Engineering, Indian Institute of Technology, Kanpur 208016, India\\}
\date{\today}
\begin{abstract}
We discuss the efficacy of evolutionary method for the purpose of structural analysis of amorphous solids. At present \textit{ab initio} molecular dynamics (MD) based melt-quench technique is used and this deterministic approach has proven to be successful to study amorphous materials. We show that a stochastic approach motivated by Darwinian evolution can also be used to simulate amorphous structures. Applying this method, in conjunction with density functional theory (DFT) based electronic, ionic and cell relaxation, we re-investigate two well known amorphous semiconductors, namely silicon and indium gallium zinc oxide (IGZO). We find that characteristic structural parameters like average bond length and bond angle are within $\sim 2\%$ to those reported by \textit{ab initio} MD calculations and experimental studies.  
\end{abstract}
\maketitle

\section{Introduction} 
\label{intro}
Computational route to design new materials with desired properties is becoming viable due to tremendous improvement of modern computers, both in terms of speed and data storage capacity.\cite{Curtarolo13,Oganov11M, Woodley08} However, reliability of computational approach to predict a material's property depends on the knowledge of it's structure. The input is derived either from experimental data or obtained via one of the crystal structure prediction techniques. While the first method is very effective for a known substance, the second route is inevitable for the development of new materials. Ideally, a crystal structure prediction technique should be capable of finding the most stable structure based on the knowledge of the chemical composition of the constituent atoms, but it remains a stiff challenge to do so. Several methods are used for this purpose; like random sampling,\cite{Karnopp63,Pickard09,Feng08} simulated annealing,\cite{Metropolis53, Kirkpatrick83, Stich91, Kresse94} minima hopping,\cite{Goedecker04, Goedecker05, Hellman07, Amsler10} meta-dynamics\cite{Parinello02} and genetic algorithms\cite{Woodley99, Woodley04} etc., although none of them are guaranteed to give the correct answer because of the complexity of exploring the multidimensional energy landscape efficiently.\cite{Wales06,Organov09} In this regard USPEX (Universal Structure Prediction: Evolutionary Xtallography), an evolutionary algorithm based method developed recently, has shown great promise so far.\cite{Glass06, Organov11, Lyakhov13} Based on first principles calculations, this method has made several startling discoveries, like the existence of a transparent phase of Na\cite{Ma09} or a superconducting phase of CaLi$_2$,\cite{Xie10} both occurring at high pressure and experimental validation has followed the computational prediction.

Despite USPEX's success in finding the crystal structure of several unknown materials,\cite{Zhu13,Zhou12,Zhang13} it has not yet been used to predict the structure of thermodynamically metastable amorphous phases. Melt-quench technique, using the \textit{ab initio} molecular dynamics (MD) simulations, remains the most preferred route for generating amorphous structures. One of the major drawbacks of MD simulation, a deterministic approach, is ``local sampling'', resulting from uneven energy landscape with multiple local minima separated by relatively high energy barriers. Moreover, in case of \textit{ab initio} MD based melt-quench technique, the resultant structure is highly likely to retain signatures of the molten phase and the method can also fail if the structure is made of complex building blocks.\cite{Drabold09} On the contrary, evolutionary algorithms are capable of ``non-local sampling'', because these stochastic methods are motivated by the natural evolution of a population based on the Darwinian hypothesis of survival of the fittest.  Because of this, evolutionary method is highly successful to predict crystal structure, which motivated us to investigate whether it can be an alternative to the \textit{ab initio} melt-quench technique for simulating amorphous structures.

In this paper, we show the effectiveness of evolutionary algorithm, as implemented in USPEX, to predict the structure of amorphous materials and to the best of our knowledge, this is the first such attempt in this direction. We choose two well known materials, amorphous form of silicon and indium gallium zinc oxide (IGZO) for this purpose and we find very good agreement by comparing our results with those obtained via \textit{ab initio} MD calculations\cite{Stich91, Durandurdu01, Noh11} and experimental studies.\cite{Fortner89, Kugler89, cho09, Hosono07}

The article is organized as follows: In Sec~\ref{sec:SeS} we present our study on amorphous Si, a mono-atomic material, followed by a discussion on amorphous IGZO, a multi-component system in Sec~\ref{sec:MeS}. We discuss various technical issues regarding convergence of the method and possibility of further structural refinement in Sec~\ref{conv} and the paper is concluded in Sec~\ref{concl}.

\section{Single-element System : Silicon}
\mylabel{sec:SeS}

\subsection{Computational Details}
\mylabel{CD}

\begin{table*}[]
\centering
\begin{tabular}{|c| c c c c c c |c|c|}
\hline
Energy Band & 1 & 2 & 3 & 4 & 5 & 6 & a-Si & c-Si\\
\hline
Semiconductor & & & & & & & &\\
ABL ({\AA}) & 2.41 &2.4  & 2.39 & 2.4 & 2.39 & 2.4 & 2.4 & 2.36 \\
ABA & 105.96$^{\circ}$ & 106.42$^{\circ}$ & 105.97$^{\circ}$ & 106.21$^{\circ}$ & 107.03$^{\circ}$ & 106.62$^{\circ}$ & 106.4 & 109.47$^{\circ}$\\
ACN (2.8 {\AA}) & 4.22 & 4.16 & 4.22 & 4.27 & 4.24 & 4.33 & 4.24 & 4\\
\hline 
Metal & & & & & & & &\\
ABL ({\AA}) & 2.61 & 2.60 & 2.59  & 2.56  & 2.56 & 2.57  & 2.58 & 2.36\\
ABA & 101.15$^{\circ}$ & 101.61$^{\circ}$ & 101.65$^{\circ}$ & 101.23$^{\circ}$ & 100.72$^{\circ}$ & 100.60$^{\circ}$ & 101.16$^{\circ}$ & 109.47$^{\circ}$\\
ACN (2.8 {\AA}) & 6.80 & 6.41 & 6.15 & 5.51 & 5.65 & 5.70 & 6.05 & 4\\
\hline
\end{tabular}
\caption{Structural parameters like average bond length (ABL), average bond angle (ABA) and average coordination number (ACN) for the semiconducting and metallic phase of a-Si, for each of the six energy bands [see Fig~\ref{fig2}]. Corresponding values for the crystalline phase are given in the last column for comparison. Averages are calculated by taking a cut-off radius of 2.8 \AA~ around every Si atom, i.e., any Si lying within this distance from a given Si atom is considered to form a bond. Note that, numerical value of coordination number depends on the choice of cut-off radius, as shown in Fig~\ref{fig3}(e) and (f). Average bond lengths and bond angles found here are within $\sim 2\%$ of the values reported for semiconducting\cite{Stich91} and metallic\cite{Durandurdu01} phase of a-Si, obtained by \textit{ab initio} MD based melt-quench technique.}
\label{t1}
\end{table*}

We use evolutionary algorithm (EA) based code USPEX \cite{Glass06, Organov11, Lyakhov13}(Universal Structure Prediction: Evolutionary Xtallography)  to sample the multidimensional configuration space. Based on the knowledge of chemical composition, this method is capable of predicting the stable structure of a particular compound for a given pressure and temperature. To begin with, we generate 24 structures by a random number generator in conjunction  with space group symmetry; each structure having 64 atoms (8 atoms) in a cubic unit cell for a-Si (c-Si) and this set constitute the \textit{first generation}.  The choice of a cell containing 64 atoms to study the amorphous phase is motivated from the previously reported \textit{ab initio} MD simulations, using a cell of similar size for this purpose.\cite{Stich91} Using first principles density functional theory (DFT)  based calculations,
\footnote{Structural relaxations are performed using density functional theory (DFT) calculations, as implemented in Quantum Espresso package,\cite{Gianozzi09} employing a plane-wave basis set with 30 Ry kinetic energy cutoff and ultra-soft pseudopotentials. Electron exchange and correlation is treated within the framework of generalized gradient approximation (GGA). Because of the large size of the supercell, we use the $\Gamma$ (0, 0, 0) point in the reciprocal space for Brillouin zone integrations. In case of fixed volume calculation, the atomic positions are relaxed until the force (energy change between two successive steps)  on each atom is less than $10^{-3}$ Ry/au ($10^{-4}$ Ry)  while the unit cell volume is kept fixed (equal to that of crystalline silicon). In case of variable volume calculation, volume of the supercell is relaxed until the pressure is less than 0.5 Kbar.} 
both fixed cell and variable cell structural relaxations are performed for all the structures. After sorting the locally optimized structures in descending order based on their energies/enthalpies (taken as the \textit{fitness parameter} depending on fixed/variable cell run), as predicted by DFT calculations, 10 structures ($\sim $ 40 $\%$)  from the top are rejected and the remaining 14 ($\sim $ 60 $\%$)  low energy structures are selected to build the \textit{next generation} by using several variational operations, like \textit{heredity} (merging sections from different pairwise combinations of structures), \textit{lattice mutation} (applying a random strain on the lattice vectors) and \textit{soft mutation} (displacing the atoms along eigenvectors of the softest phonon modes). All the operations are performed in such a way that the total number of atoms in the unit cell is conserved. Among the 24 of the \textit{next generation} structures, 19 ($\sim 80 \%$)  of them are generated by variational operations (using the 14 low energy structures obtained in the \textit{previous generation}) and the remaining 5 structures are generated by random sampling. In addition to that, 2 of the lowest energy structures from the previous generation are always kept in the next generation. We let the simulation continue for 20 generations, which gives us a huge database ($\sim 450$ structures)  for further statistical analysis.  

\subsection{Crystalline silicon (c-Si)}
\begin{figure}[]
\begin{center}
\includegraphics[width=0.7 \linewidth]{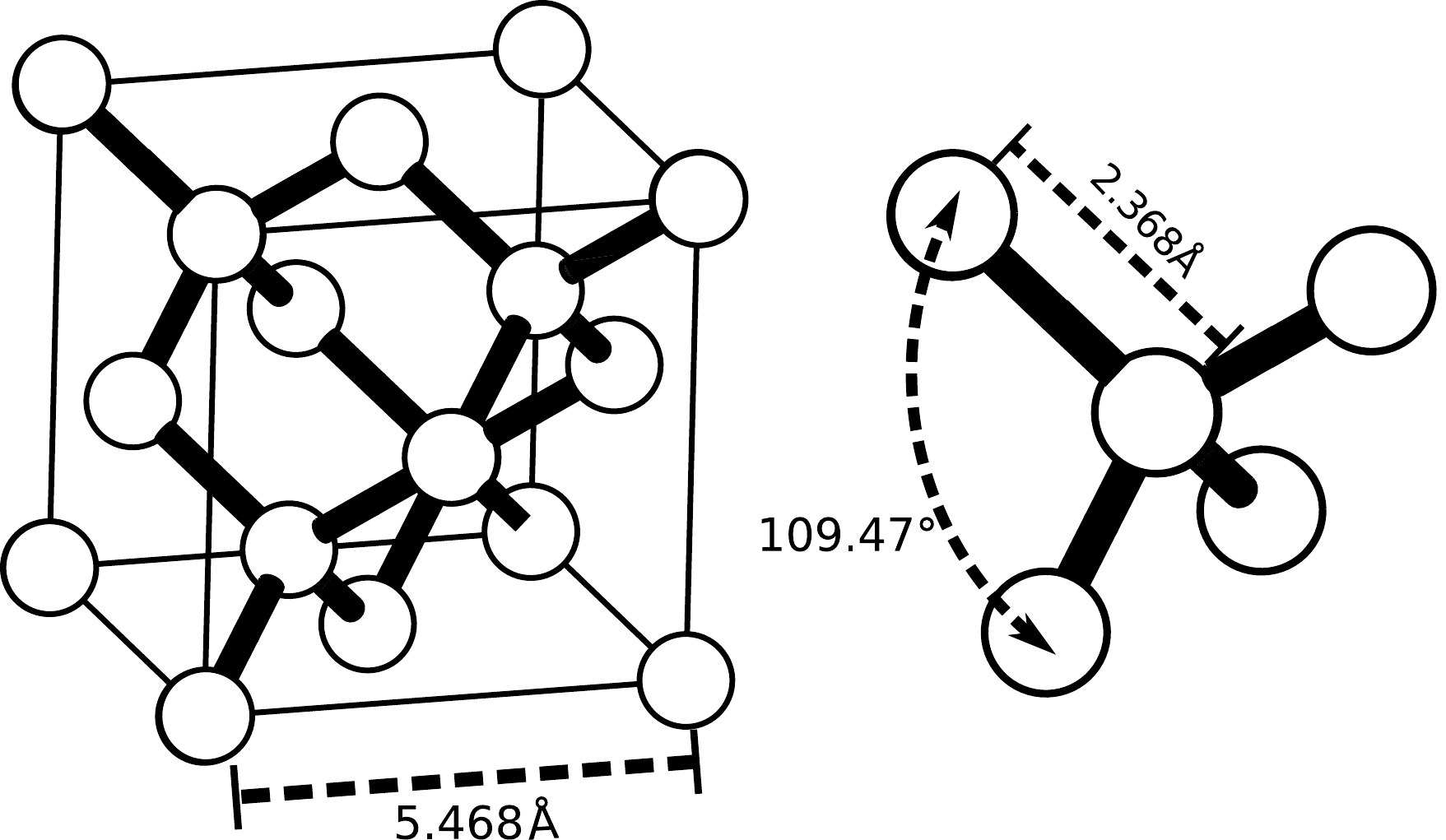}
\caption{Unit cell of crystalline silicon obtained from ground state search using USPEX and visualized using Xcrysden.\cite{Kokalj99, Kokalj03}}
\label{fig1}
\end{center}
\end{figure}
In order to validate the pseudopotential and other computational parameters used in this work, initially we calculate the structural and electronic properties of crystalline silicon. A ground search is conducted using USPEX, with the number of Si atoms in the unit cell constrained to be 8. As expected, crystalline diamond cubic phase of silicon with a space group of Fd-3m (227) [see Fig~\ref{fig1}(a)] is found to be the lowest energy structure. The calculated value of equilibrium lattice parameter of this phase is equal to 5.47 {\AA}, which is in good agreement with experimentally observed value of 5.43 {\AA}.\cite{Okada84} Each of the silicon atoms in the crystal are four-coordinated with a bond angle of 109.47$^{\circ}$. The first, second and third nearest neighbor distances are found to be 2.37 {\AA}, 3.87 {\AA} and 4.53 {\AA}, respectively. In terms of electronic property, Si is confirmed to be an indirect band gap semiconductor, but the calculated value of band gap (0.7 eV) is found to be considerably less than the actual magnitude of 1.1 eV.\cite{Chelikowsky74} This underestimation is a well known limitation of DFT to predict the band gap correctly, which can be improved by GW approximation\cite{Hott91} to give a better match with experimental measurement. Since the above results are in good agreement with the computational and experimental data available in the literature for c-Si, we use the same set of computational parameters for a-Si simulations.

\subsection{Amorphous silicon (a-Si)}       
\begin{figure}[]
\begin{center}
\includegraphics[width=0.85 \linewidth]{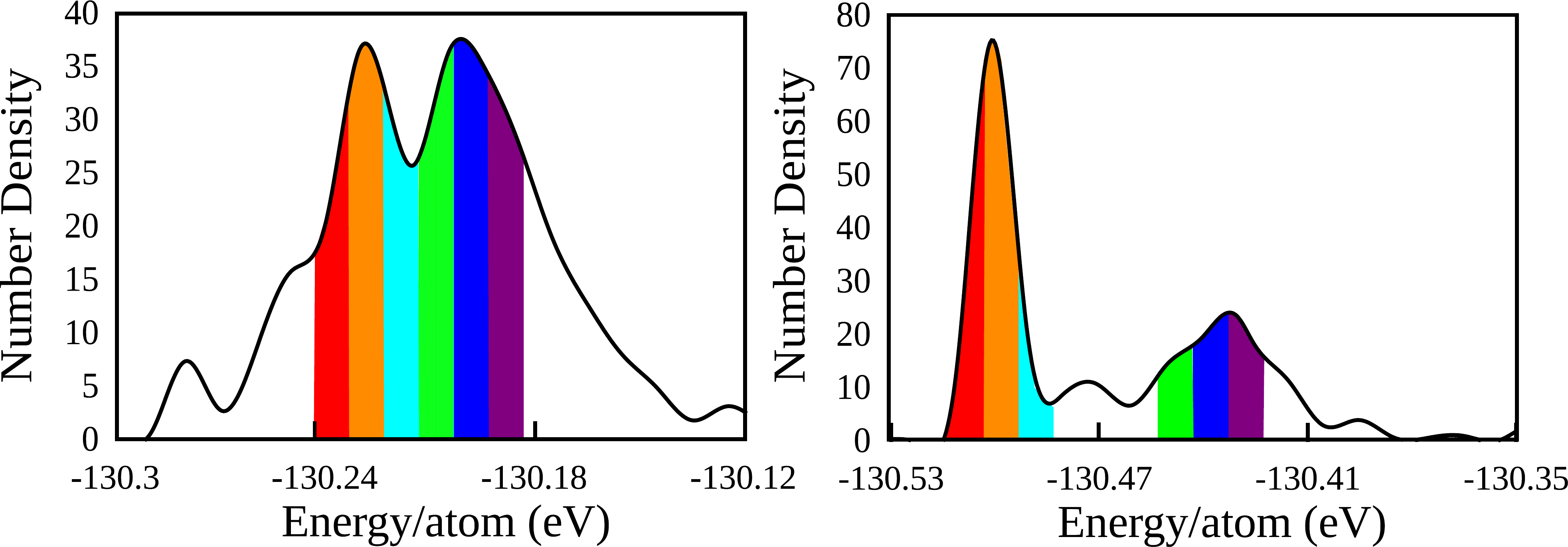}
\caption{Number density of structures (NDOS) in the energy landscape of a-Si: (a) semi-conducting and (b) metallic phase. Each column represents number of structures within 10 meV energy range.}
\label{fig2}
\end{center}
\end{figure}

\begin{figure*}[]
\begin{center}
\vspace{0.2in}
\includegraphics[width=0.85 \linewidth]{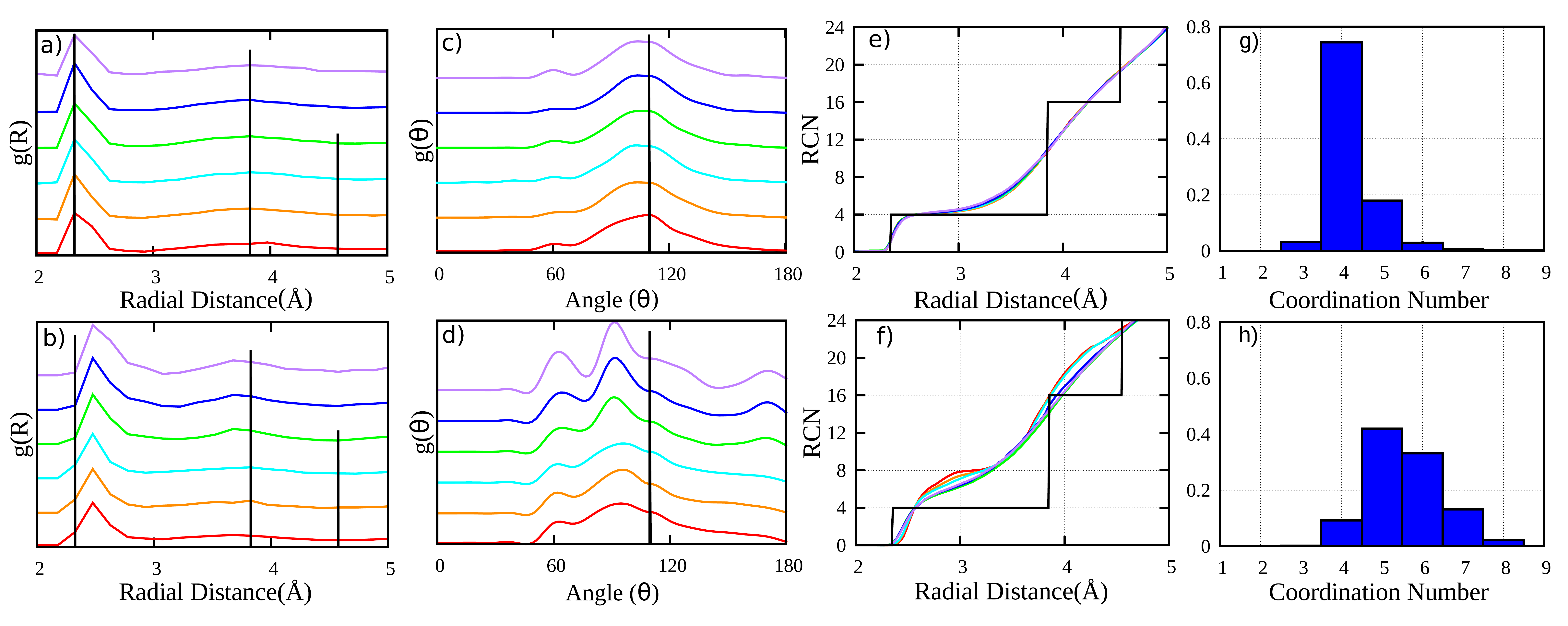}
\caption{Plots showing structural analysis of a-Si. Radial distribution function (RDF), bond angle distribution function (BADF), running coordination number (RCN) and distribution of coordination number for semiconducting a-Si (first row), metallic a-Si (second row) are plotted in first, second, third and fourth column, respectively. Colors used here are in accordance with the colors of the columns denoting the energy bands in Fig~\ref{fig2}. For c-Si, delta function like peaks in RDF and BADF and sharp steps in RCN are shown in black lines [consult Fig~\ref{fig1} and Table~\ref{t1}]. Coordination number distribution, calculated using Si-Si cutoff distance of 2.8 \AA, clearly shows significant number of over-coordinated Si atoms in the metallic phase, while majority of Si atoms are four coordinated in case of semiconducting phase of a-Si, same as c-Si.}
\label{fig3}
\end{center}
\end{figure*}

As stated in Sec~\ref{CD}, we sample the configuration space starting with a set of 24 structures, each having 64 atoms arranged in a cubic unit cell. Structural relaxations are performed in two different routes; (a) fixed volume, maintaining the cubic symmetry and (b) variable volume, allowing the lattice vectors to change. Interestingly, we find that the former and latter route yields a-Si structures, which are semiconducting and metallic, respectively. Consecutive generations are constructed using USPEX, as described in Sec~\ref{CD}, followed by structural relaxation. After repeating this exercise for 20 generations, we build a database of $\sim$450 structures and sort them according to total energy per atom. Then we calculate the number density of structures (NDOS), defined as the number of structures within 5 meV energy range. The plot of NDOS for semiconducting and metallic phase is shown in Fig~\ref{fig2}(a) and (b), respectively. Note that, certain regions of NDOS plots have sharp peaks due to the existence of highly degenerate amorphous structures. We select such regions and divide the structures in six different energy bands of width 10 meV each [see the colored columns in Fig~\ref{fig2}], such that each band consists of at least 50 structures or more. We exclude the structures from the region of low NDOS, because their relative frequencies are very small, making them highly improbable than compared to the structures selected from a region of high NDOS. In this process, we include only $\sim$75-80\% of the structures from our original database of $\sim$450 for further statistical analysis and ignore the rest. Based on radial distribution function (RDF) and bond angle distribution function (BADF) plots [Fig~\ref{fig3}], we conclude that the predicted structures are amorphous. While RDF and BADF characteristics is going to be discussed in detail later, we point out that only the first peak in RDF is relatively prominent [see Fig~\ref{fig3}(a) and (b)] because amorphous materials are merely having short range order. Lack of long range order is evident from the absence of other peaks (occouring at higher distance than compared to the bond length) in RDF and the BADF plots showing wide range for bond angle distribution [see Fig~\ref{fig3}(c) and (d)].  

 \begin{figure}[]
\begin{center}
\includegraphics[width=0.95 \linewidth]{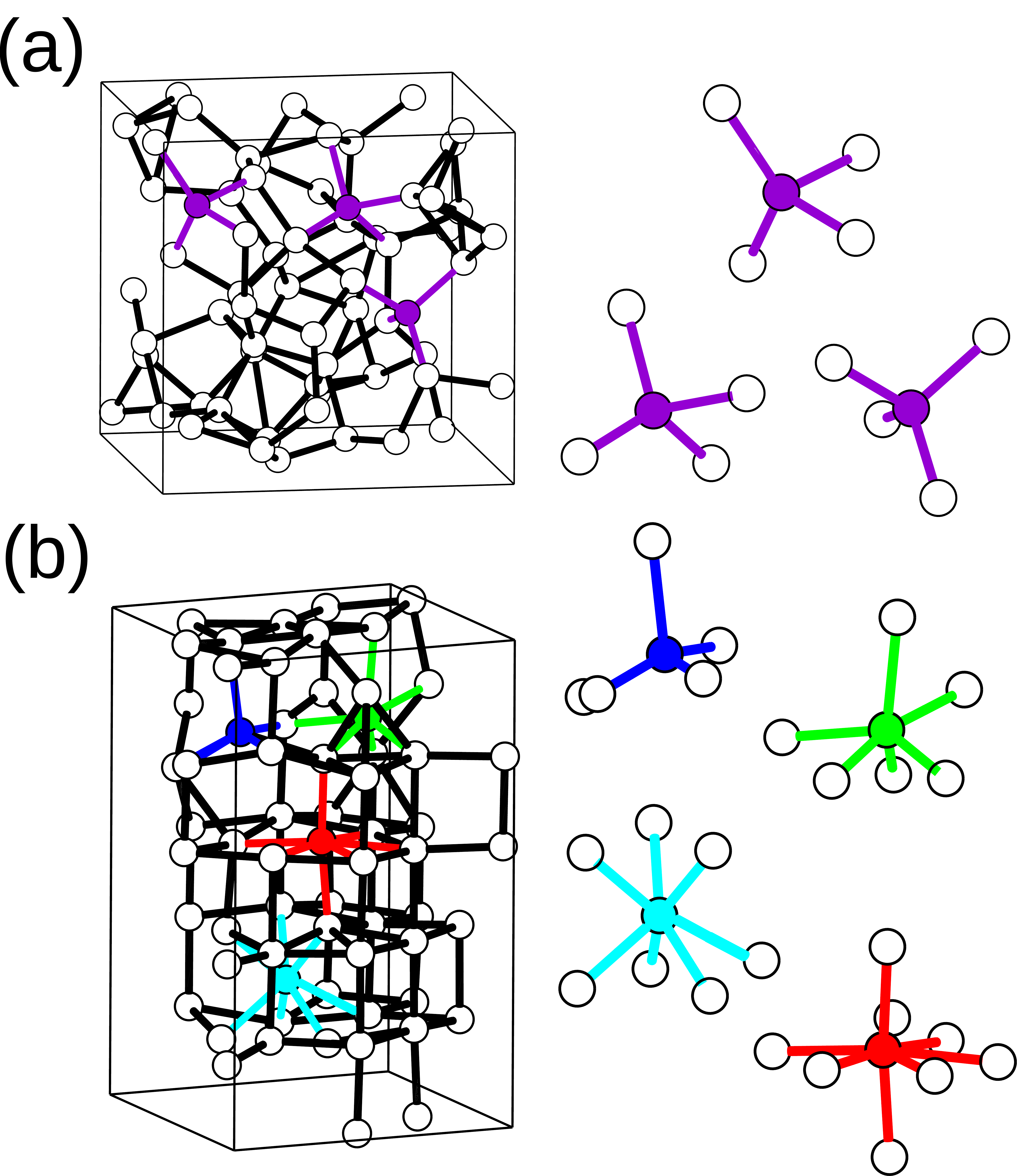}
\caption{Atomic arrangement of (a) semiconducting and (b) metallic phase of a-Si. Atoms with different coordination number are denoted by different colors ; 4: violet, 5: blue, 6: green , 7: cyan and 8: red).}	
\label{fig4}
\end{center}
\end{figure}

Average bond length (ABL), average bond angle (ABA) and average coordination number (ACN) are calculated by averaging the respective quantities over all the structures belonging to a particular band and reported in Table~\ref{t1} for each of the six energy bands. Since the energy difference among different bands are relatively small, it is not surprising that they have very similar values of ABL, ABA and ACN. Further averaging over the numbers obtained for six different energy bands, we find the value of ABL, ABA and ACN to be equal to 2.4 \AA, 106.4$^{\circ}$ and 4.24 for the semiconducting phase and 2.58 \AA, 101.16$^{\circ}$ and 6.05 for the metallic phase of a-Si. The structural parameters given in Table~\ref{t1} are found to be within 2\% of the values reported for both semiconducting~\cite{Stich91} and metallic phase~\cite{Durandurdu01} of a-Si, obtained via melt-quench technique using \textit {ab initio} molecular dynamics (MD) simulations. This clearly shows the effectiveness of evolutionary approach for predicting amorphous structures.


Encouraged by the success to generate amorphous structures, we perform detailed analysis in terms of two and three body correlation function like radial distribution function (RDF) and bond angle distribution function (BADF). RDF is defined as the number of atoms in a spherical shell divided by the volume of the shell and number density of atoms in the system. We further calculate the \textit{mean} RDF by taking average over all the structures belonging to a particular energy band and show the plots for each of the six bands in Fig~\ref{fig3} for both (a) semiconducting and (b) metallic phase. Because of translation symmetry (long range order), RDF for a crystalline material has sharp peaks at well defined intervals coming from first, second and consecutive neighbors, as shown by black lines in Fig~\ref{fig3}(a) and (b). Comparing with RDF of c-Si, it is found that the first peak is very prominent but other peaks are broadened considerably in case of a-Si. As mentioned before, this is the classic signature of amorphous materials, having only short range order manifested by sharp initial peak in RDF, but higher order peaks are missing due to lack of periodicity. Note that the mean radial distance  of the first neighbor peak of the semiconducting phase coincides with that of crystalline silicon at $\sim$ 2.4 {\AA} [Fig~\ref{fig3}(a)]. On the contrary, the first neighbor peak of the metallic phase shifts towards higher radial distance [Fig~\ref{fig3}(b)] of $\sim$ 2.5 {\AA}. Again, our results are in good agreement with those simulated via \textit{ab initio} MD route\cite{Stich91, Durandurdu01} and experimental studies.\cite{Fortner89, Kugler89}

Bond angle is calculated by measuring the angle ($\theta$) between two vectors, originating from a reference atom and connecting two of it's neighbors. The neighbor list includes all the atoms within a cut-off radius of 2.8 {\AA} from the reference atom. Unlike c-Si, having unique bond angle of 109.5$^{\circ}$ resulting from sp$^3$ hybridization, $\theta$ values are distributed over a large range for amorphous structures. We further calculate the \textit{mean} BADF by averaging the bond angle distribution of all the structures belonging to a particular energy band; as shown in Fig~\ref{fig3}(c) (semiconducting a-Si) and (d) (metallic a-Si) for each of the six bands.  In case of the semiconducting phase, a broad peak is observed in the range 60--120$^{\circ}$, with a mean bond angle ($\sim$106--107$^{\circ}$, see Table~\ref{t1}), which is very close to the bond angle of c-Si, shown by the delta function like peak at $\theta=109.5^{\circ}$. On the other hand, two peaks are observed about 60$^{\circ}$ and 90$^{\circ}$, followed by a small hump from 150--170$^{\circ}$, for the metallic phase of a-Si. As a result, mean bond angle ($\sim$100--101$^{\circ}$, see Table~\ref{t1}) deviates significantly from it's value for c-Si. Not only the qualitative features, like a single peak for semiconducting and dual peaks for metallic a-Si, but also the quantities like location of peaks and their widths in BADF plots are consistent with those obtained via \textit{ab initio} MD simulations.~\cite{Stich91, Durandurdu01} 

From the data presented in Table~\ref{t1}, clearly the semiconducting phase of a-Si is more closely related to the c-Si than compared to the metallic phase of a-Si, in terms of structural parameters like bond length, bond angle and coordination number. Further analysis is performed in terms of running coordination number (RCN) plots, showing the average CN as a function of radial distance from a central atom. The \textit{mean} value of RCN is calculated for each of the six energy bands by averaging the quantity over all the structures belonging to a particular band and illustrated in Fig~\ref{fig3}(e) and (f) for the semiconducting and metallic phase of a-Si, respectively. Note that the sharp steps observed in case of c-Si are absent for a-Si, which is due to lack of long range order in the latter. While a plateau region is observed for the semiconducting phase, extending from $\sim$2.4 to 3 \AA~ radial distance where the CN is nearly constant ($\sim$4), but for the metallic phase CN is found to rise continuously in the stated range, attaining a value of $\sim 7-8$ at 3 \AA~ radial distance. The difference between the semiconducting and metallic phase highlighted in RCN plots suggests that, similar to c-Si, four-coordinated silicon atoms (in the form of distorted tetrahedrons) remain the basic building blocks in semiconducting a-Si. On the other hand, absence of a plateau in the RCN plot for the metallic phase of a-Si indicates the presence of silicon atoms with more than one type of coordination.

\begin{table*}[ht]
\centering
\begin{tabular}{|c| c c c c c c |c|c|}
\hline
Energy Band & 1 & 2 & 3 & 4 & 5 & 6 & a-IGZO & c-IGZO\\
\hline
ABL ({\AA}) & & & & & & & &\\
\hline 
In-O & 2.23 & 2.24 & 2.23 & 2.23 & 2.25 & 2.24 & 2.24 & 2.2 \\
Ga-O & 1.96 & 1.96 & 1.97 & 1.97 & 1.99 & 1.97 & 1.97 & 1.93\\ 
Zn-O & 2.07 & 2.08 & 2.08 & 2.08 & 2.09 & 2.09 & 2.08 & 2 (2.54)\\
\hline
ABA & & & & & & & &\\
\hline
O-In-O & 105.58$^{\circ}$ & 105.04$^{\circ}$ & 105.11$^{\circ}$ & 105.00$^{\circ}$ & 104.26$^{\circ}$ & 104.79$^{\circ}$ & 104.8$^{\circ}$ & 81$^{\circ}$, 99$^{\circ}$ \\ 
O-Ga-O & 107.79$^{\circ}$ & 107.58$^{\circ}$ & 107.64$^{\circ}$ & 107.62$^{\circ}$ & 106.36$^{\circ}$ & 107.40$^{\circ}$ & 107.4$^{\circ}$ & 90$^{\circ}$, 120$^{\circ}$ \\
O-Zn-O & 108.96$^{\circ}$ & 108.42$^{\circ}$ & 108.47$^{\circ}$ & 108.03$^{\circ}$ & 107.37$^{\circ}$ & 107.71$^{\circ}$ & 108.16$^{\circ}$ & 74$^{\circ}$, 106$^{\circ}$, 112$^{\circ}$\\
\hline
ACN (3.00 {\AA}) &  &  &  &  &  & & &\\
\hline
In-O & 4.79 & 4.83 & 4.87 & 4.84 & 4.65 & 4.82 & 4.8 & 6\\
Ga-O & 4.39 & 4.4  & 4.44 & 4.42 & 4.30 & 4.36 & 4.39 & 5\\
Zn-O & 4.22 & 4.22 & 4.33 & 4.37 & 4.35 & 4.41 & 4.32 & 5\\

\hline
\end{tabular}
\caption{Average bond length, bond angle and coordination number for a-IGZO and c-IGZO. Averages are calculated by taking a cut-off radius of 3 \AA~ around every metal, i.e., any atom lying within this distance from a particular metal atom is considered to be bonded to that metal. Corresponding values for the crystalline phase are given in the last column for comparison. Note that, numerical value of coordination number depends on the choice of cut-off radius, as shown in Fig~\ref{fig7}(g), (h) and (i). Average bond lengths found here are within $\sim 2\%$ of the experimentally observed\cite{cho09} values.}
\label{t3}
\end{table*}

This is further analyzed in terms of coordination number distribution of the two amorphous phases, as shown in Fig~\ref{fig3}(g) and (h). The calculation includes all the structures lying within the NDOS peaks marked in Fig~\ref{fig2} and we set the Si-Si cut-off distance equal to 2.8 \AA~ to get the coordination number distribution. Evidently, significant number of Si atoms in the metallic a-Si phase have higher coordination than compared to the ideal value of four, observed in c-Si. On the other hand, semiconducting phase of a-Si has very similar coordination characteristics as found in c-Si.
This is further illustrated  in Fig~\ref{fig4}, prepared by Xcrysden visualization tool,\cite{Kokalj99, Kokalj03} where each of the unit cell of semiconducting and metallic a-Si contains 64 Si atoms.  As shown in the diagram, the semiconducting phase contains mostly four-coordinated atoms, while the metallic phase consists of atoms which are five to eight coordinated. Because of this, the metallic phase ($\sim$ 3 g/cm$^3$) has higher density than compared to that of the semiconducting phase ($\sim$ 2.3 g/cm$^3$), which can be the simplest way of distinguishing the two phases experimentally. 
       
\section{Multiple-element System : Indium Gallium Zinc Oxide}
\mylabel{sec:MeS}

\subsection{Crystalline IGZO (c-IGZO)}
\label{cigzo}

\begin{figure}[]
\begin{center}
\includegraphics[width=0.5 \linewidth]{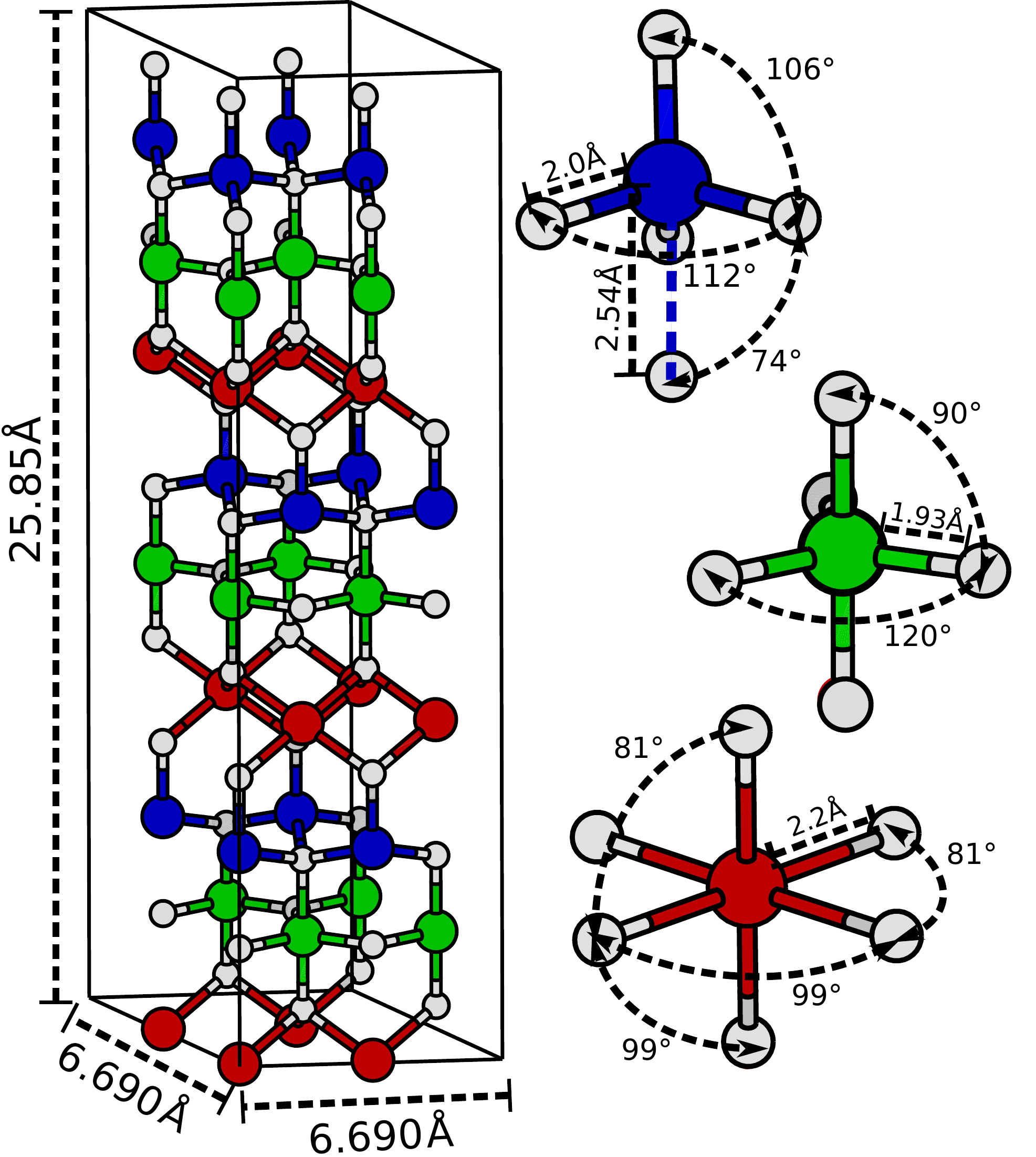}
\caption{Unit cell of crystalline IGZO, repeated $2\times 2$ along the a and b axis, lying in the horizontal plane. This is adopted from the work of \citeauthor{Meux15}\cite{Meux15} In, Ga, Zn and O atoms are shown in red, green, blue and gray, respectively.}
\label{fig5}
\end{center}
\end{figure}

Following it's effectiveness for structure prediction of a-Si, we try to extend it for a multi-element system, indium gallium zinc oxide (InGaZnO$_4$). In order to validate the pseudopotential and other calculation parameters, initially we calculate the structure and electronic band structure of c-IGZO and compare our result with the values reported in the literature. Crystalline form of IGZO is composed of alternate layers of InO$_2^-$ and GaZnO$_2^+$ and each unit cell is formed by six such layers stacked along the vertical direction. As shown in Fig~\ref{fig5}, the unit cell contains 21 atoms in total [chemical formula In$_3$Ga$_3$Zn$_3$O$_{12}$], where the In atoms occupy the octahedral sites [coordination number 6] while both the Zn and Ga atoms  occupy the trigonal-bipyramidal sites [coordination number 5].  
The equilibrium lattice parameters after relaxing the structure are found to be a = b = 3.345 \AA~ and c = 25.85 \AA~ and the bond lengths of In, Ga and Zn with oxygen are calculated to be 2.2 \AA, 1.93 \AA~ and 2 \AA~ [2.54 \AA], respectively. Note that, one of the two out of plane Zn-O bonds is 0.54 \AA~ longer than the other. The O-In-O bond-angle deviates by $\pm 9^\circ$ from it's ideal value of 90$^\circ$ in  a perfect octahedron. Although both Ga and Zn occupies the trigonal-bipyramidal sites, corresponding bond-angles differ significantly. In particular, O-Ga-O bond-angles are equal to $90^\circ$ and $120^\circ$ for out-of-plane and in-plane bonds, respectively, matching exactly to the values for a perfect trigonal-bipyramid. On the other hand, O-Zn-O bond-angles differ significantly from their ideal values and they are measured to be equal to $112^\circ$ for in-plane and $74^\circ$ and $106^\circ$ for out-of-plane bonds. However, these values can not be compared directly with experimental results, because, as reported in the literature, Ga and Zn atoms randomly occupy the trigonal-bipyramidal sites in GaZnO$_2^+$ layer.\cite{Orita00, Omura09, Chen11, Meux15, Hosono07} Thus, it will be more appropriate to calculate the average of bond-lengths and angles for in-plane and out-of-plane bonds for Ga and Zn polyhedron and compare them with experimental data. Using this approach, we find the in-plane bond-length and angle to be equal to 1.97 \AA~ and 116$^\circ$ in GaZnO$_2^+$ layer. The out-of-plane bond-lengths are equal to 1.97 \AA~ and 2.24 \AA~, while the corresponding bond-angles are 98$^\circ$ and 82$^\circ$.  The structural parameters obtained are in good  agreement with both calculated value and experimental data.\cite{Orita00, Omura09, Chen11, Meux15, Hosono07} We also confirm that c-IGZO is a direct band gap semiconductor, with both valence band maximum (VBM) and conduction band minimum (CBM) being located at the $\Gamma$ point. The calculated band gap is $\sim$ 1.1 eV, which is in good agreement with earlier predictions based on DFT calculations, but significantly underestimated in comparison with experimental results (3.3 eV),\cite{Orita00, Meux15, Chen11} and this can be improved by GW approximation.\cite{Hott91} Thus, overall agreement of our simulation results of c-IGZO with both theoretical and experimental data available in the literature confirms the validity of the pseudopotential and computational parameters and we use the same for a-IGZO simulations.


\subsection{Computational details}

In addition to the variational operators mentioned in Section~\ref{CD}, we also use \textit{permutation} operation for multi-element system. The permutation operation essentially exchanges the position of atoms belonging to different elements. It considerably improves the diversity of the structures obtained and is of great significance while treating multiple-element systems. Apart from this, the simulation procedure followed here is same as before. The number of generations considered is 16, with each generation comprising of 32 structures, each made up of unit cells containing 84 atoms (In$_{12}$Ga$_{12}$Zn$_{12}$O$_{48}$). A database of $\sim$ 500 structures is generated and used for further analysis.  

\subsection{Amorphous IGZO (a-IGZO)}

\begin{figure}[]
\begin{center}
\includegraphics[width=0.8 \linewidth]{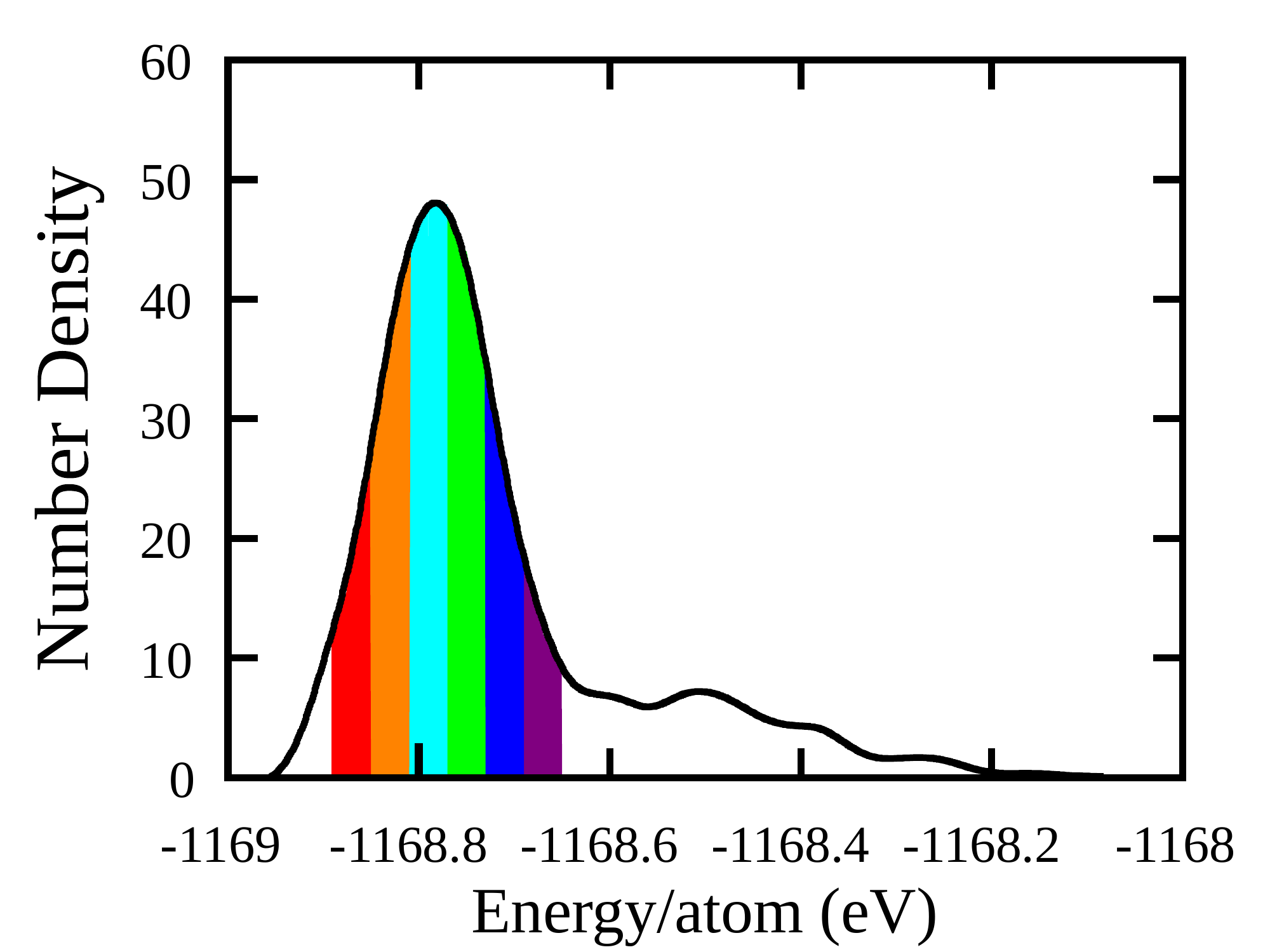}
\caption{Number density of structures (NDOS) in the energy landscape of a-IGZO. Each column represents number of structures within 40 meV energy range.}
\label{fig6}
\end{center}
\end{figure}

The NDOS peak, which includes $\sim$80\% of the 496 structures present in the original database, is divided in six energy bands of width 40 meV [see Fig~\ref{fig6}] and each of them consisting of at least 50 structures or more. Considering all the structures present in a band, average values of bond length, bond angle and coordination number are calculated and reported for each of the six energy bands in Table~\ref{t3}. Since the energy difference among different bands are relatively small, it is not surprising that they have very similar values of ABL, ABA and ACN. Further averaging over the numbers obtained for six different energy bands, we find the value of ABL, ABA and ACN to be equal to 2.24 \AA, 104.8$^\circ$ and 4.8 for In-O, 1.97 \AA, 107.4$^\circ$ and 4.39 for Ga-O and 2.08 \AA, 108.16$^\circ$ and 4.32 for Zn-O polyhedra. Calculated bond-lengths are in good agreement with the experimentally measured values of 2.3 {\AA}, 1.96 {\AA} and 2 {\AA} for In-O, Ga-O and Zn-O, respectively.\footnote{\citeauthor{cho09}\cite{cho09} reported two bond lengths, measuring 1.91 {\AA} and 2.1 {\AA}, for Zn-O and the average is matching with calculated value of $\sim 2$ \AA} Coordination numbers are slightly under-estimated when compared to an experimental study by \citeauthor{cho09},\cite{cho09} but by increasing the cutoff by 0.2 \AA, we can get a better match with the values reported in their paper. 

    
\begin{figure*}[]
\begin{center}
\includegraphics[width=0.85 \linewidth]{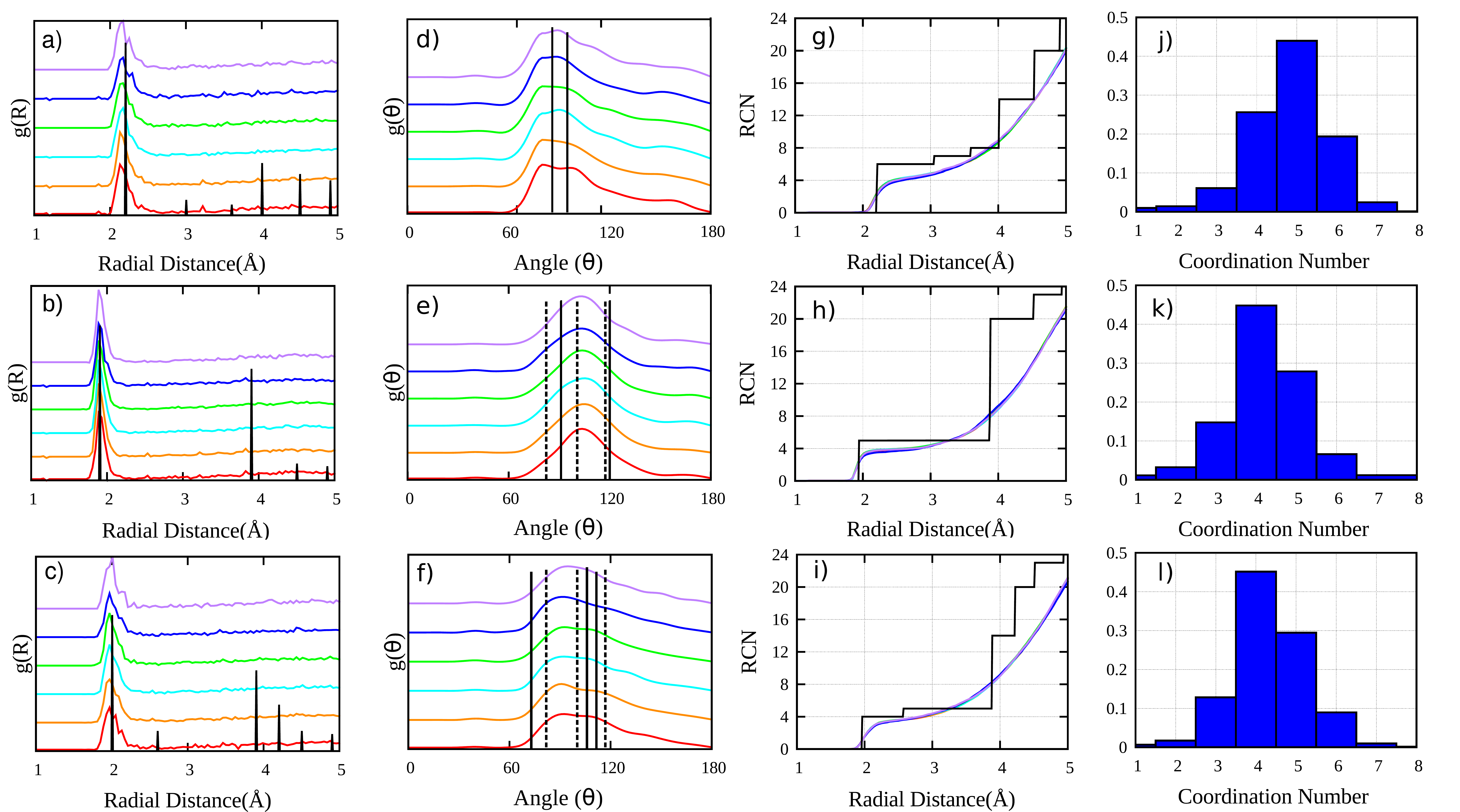}
\caption{Plots showing structural analysis of a-IGZO. RDF, BADF, RCN and distribution of coordination number for In-O (first row), Ga-O (second row) and Zn-O (third row) is plotted in first, second, third and fourth columns, respectively. Colors used in RDF, BADF and RCN plots are in accordance with the energy bands denoted in Fig~\ref{fig6}. For c-IGZO, delta function like peaks in RDF and BADF and sharp steps in RCN are shown in black lines [consult Fig~\ref{fig5} and Table~\ref{t3}]. As discussed in the text, for Zn and Ga in c-IGZO, it will be more appropriate to compare the average in-plane and out-of-plane bond-angles in MO$_5$ polyhedron with experimental results and they are shown by dashed line in BADF plots. Coordination number distribution, calculated using metal-oxygen cutoff distance of 3 \AA, clearly shows significant number of under-coordinated metal atoms in a-IGZO.}
\label{fig7}
\end{center}
\end{figure*}

\begin{figure}[]
\begin{center}
\includegraphics[width=0.6 \linewidth]{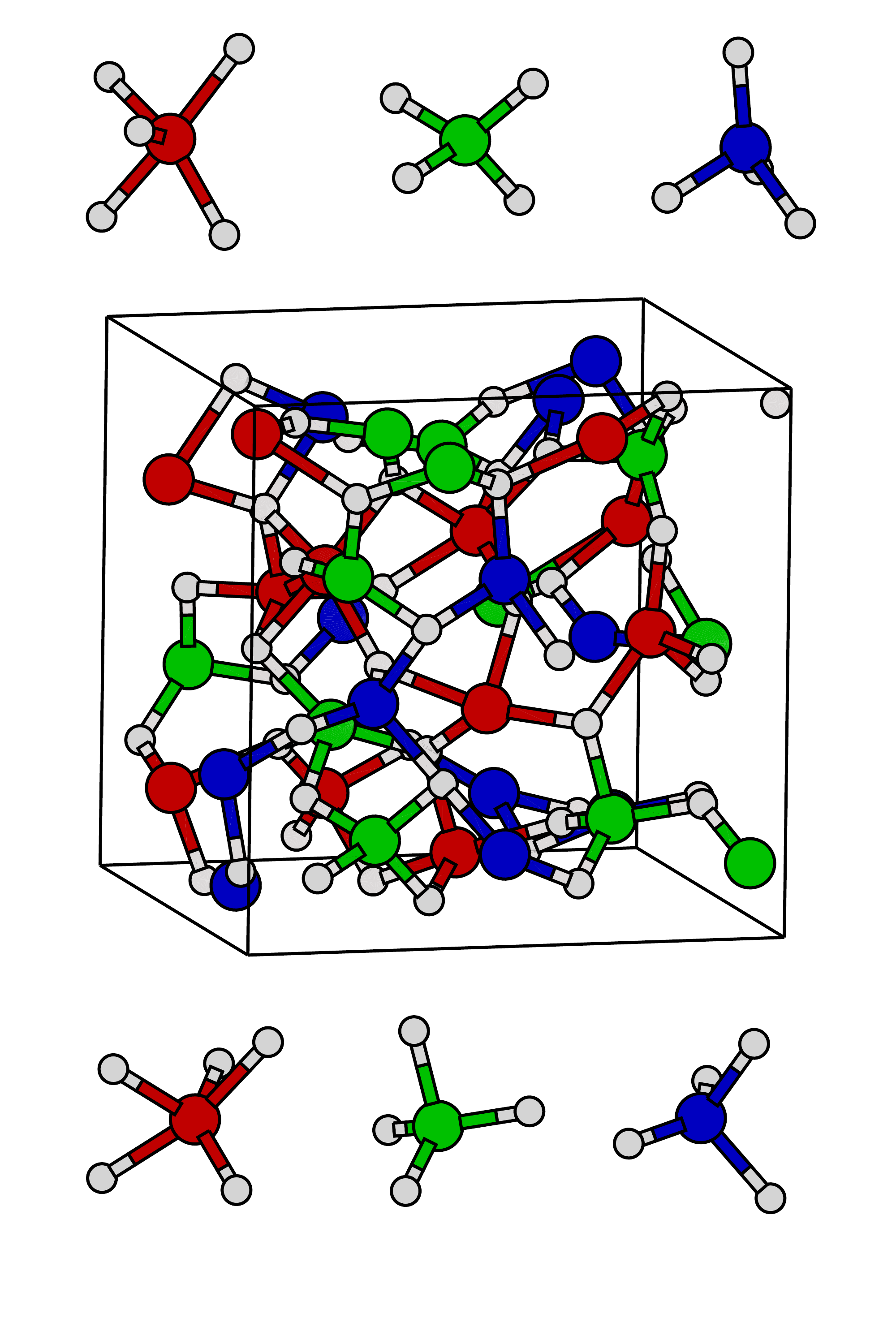}
\caption{Atomic arrangement of a representative unit cell of a-IGZO. Colors used for constituent atoms are same as Fig~\ref{fig5}. Some of the under-coordinated metal atoms are shown. Most importantly, metal-metal or oxygen-oxygen bonds are absent, as expected.}
\label{fig8}
\end{center}
\end{figure}

Further analysis is performed in terms of RDF, BADF and RCN by calculating their \textit{mean} values (averaging done over all the structures belonging to a particular energy band), as shown in Fig~\ref{fig7}(a)-(i). Since c-IGZO consists of three cations, each forming a polyhedron with oxygen atoms located at the vertexes [see Section~\ref{cigzo}], the correlation functions and RCN are calculated for three pairs, In-O, Ga-O and Zn-O. Note that, in all three of the RDF plots the first peak is significantly broadened than compared to the sharp peak observed in crystalline material and the higher order peaks are absent, which is a typical characteristic of amorphous solids, originating from the lack of long range order.

The radial distribution of oxygen with respect to In [see Fig~\ref{fig7}(a)] shows a broadened peak, with its maximum at a radial distance approximately equal to the average bond length reported in Table~\ref{t3}. This is further corroborated by the RCN plot [see Fig~\ref{fig7}(g)], where a rapid increase of coordination number is observed around $\sim$ 2.2 {\AA}. Due to lack of long range order, value of CN increases continuously with radial distance in a-IGZO and the steps observed in case of c-IGZO are absent. The radial distribution of oxygen with respect to Ga and Zn are illustrated in Fig~\ref{fig7}(b) and (c), respectively, both showing features similar to amorphous solids. As shown in the figures, the RDF of oxygen with respect to both Ga and Zn has a broadened peak at a radial distance of $\sim$ 2\AA, which is comparable to the average bond length, as reported in Table~\ref{t3}. This is in good agreement with the RCN plots [see Fig~\ref{fig7}(h) and (i)], where a sharp rise of CN is observed around $\sim$ 2 {\AA}. Since the first peak of RDF represents the metal-oxygen bond-length,\footnote{Note one small exception for Zn in the crystalline phase, where the small second peak also represents a metal-oxygen bond, identified as the slightly longer (2.54 \AA) out of plane Zn-O bond.}
based on the plots we can conclude that the bond-lengths are similar in crystalline and amorphous phase. 

Relative orientation among the constituent atoms are analyzed in terms of bond-angles measured for O-In-O [Fig~\ref{fig7}(d)], O-Ga-O [Fig~\ref{fig7}(e)] and O-Zn-O [Fig~\ref{fig7}(f)] triplets. A large distribution of bond-angles, peaked roughly around the average bond-angle [reported in Table~\ref{t3}] is a typical signature of amorphous material. Corresponding bond-angles for the crystalline structure are marked by sharp delta function like peaks drawn using solid lines. As discussed before, since it is more appropriate to represent average in-plane and out-of-plane bond angles in MO$_5$ [M=Ga/Zn] polyhedron in case of c-IGZO, we also show them using dashed lines in Fig~\ref{fig7}(e) and (f). 

Coordination number distribution is analyzed in Fig~\ref{fig7}(g), (h) and (i) for In-O, Ga-O and Zn-O, respectively. The calculation includes all the structures lying within the NDOS peak [see Fig~\ref{fig6}] and we set the metal-oxygen cut-off distance equal to 3 \AA~ to get the coordination number distribution. Evidently, significant number of metal atoms in a-IGZO are under-coordinated than compared to the ideal value of 6 for In and 5 for Ga/Zn, observed in c-IGZO. This explains the relatively lower value of average coordination number for metal atoms in a-IGZO than that of c-IGZO, as reported in Table~\ref{t3}. This can also be confirmed by visualizing [using Xcrysden~\cite{Kokalj99, Kokalj03}] a representative unit cell of In$_{12}$Ga$_{12}$Zn$_{12}$O$_{48}$, illustrated in Fig~\ref{fig8}. The figure also confirms that metal-metal or oxygen-oxygen bonds are non-existent, as expected. 

\begin{figure}[]
\centering
\includegraphics[width=0.75 \linewidth]{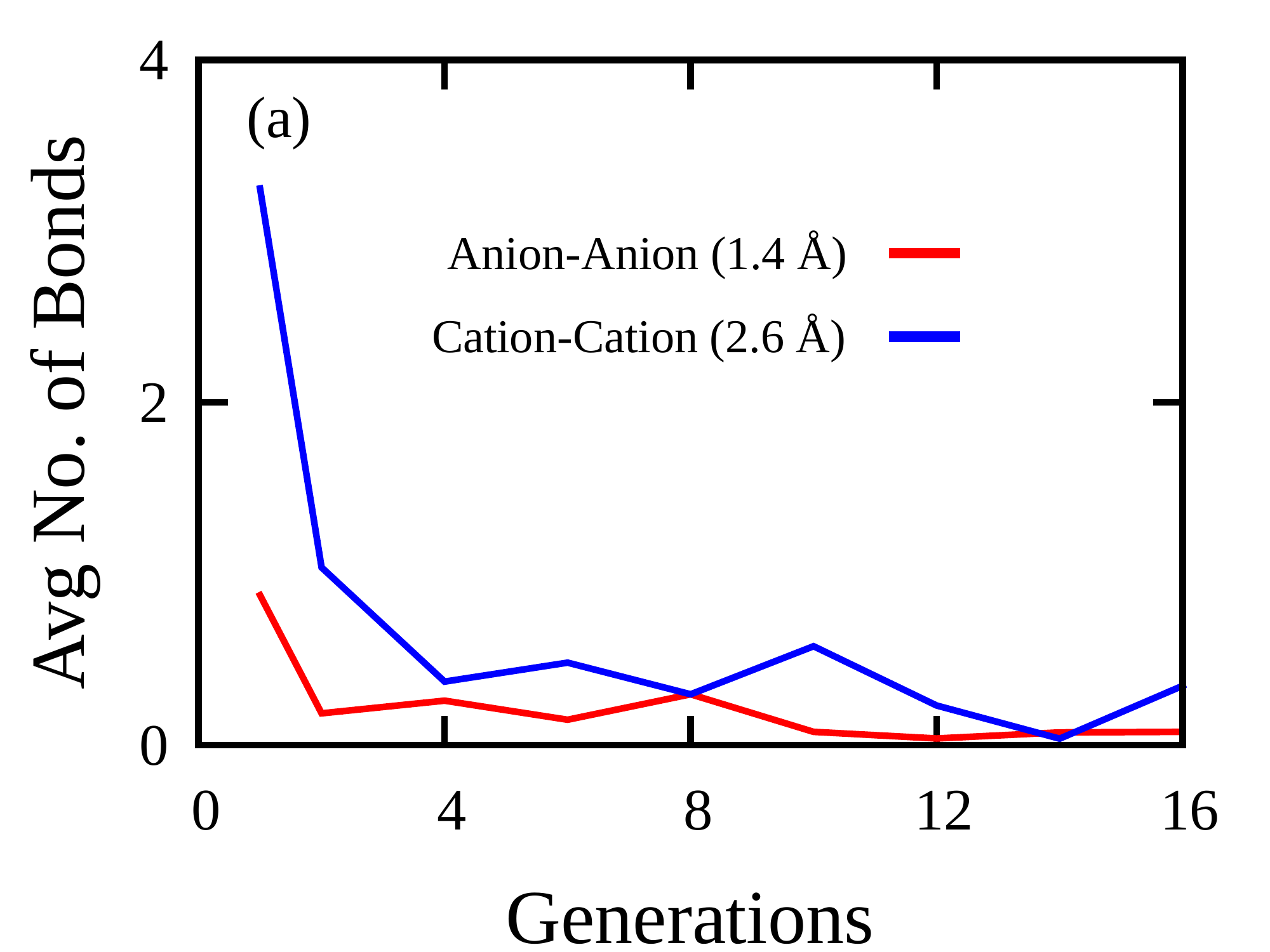}
\includegraphics[width=0.85 \linewidth]{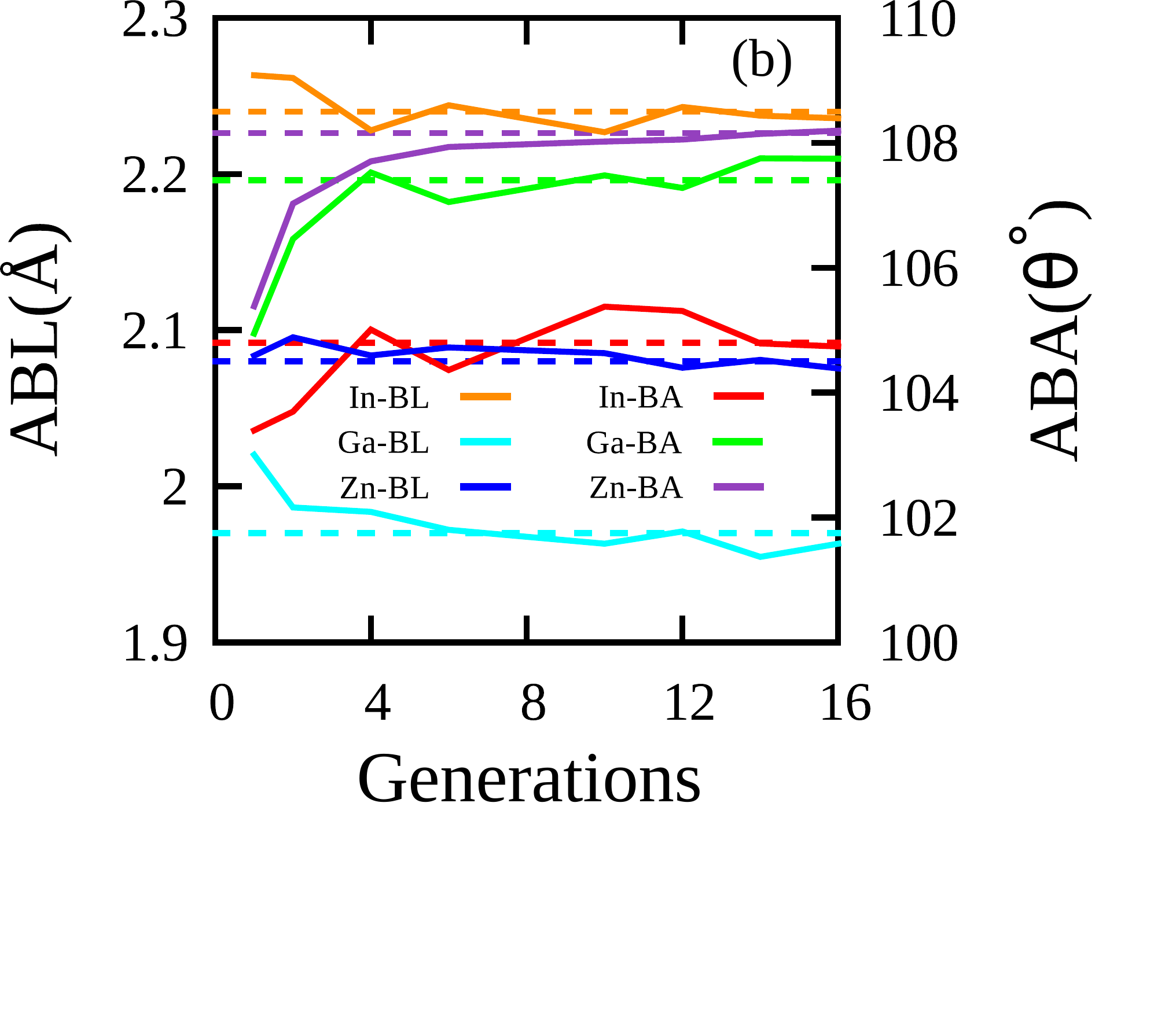}
\caption{Plots showing (a) number of cation-cation and anion-anion bonds, which are ``undesirable'' in a-IGZO structure and (b) ABL and ABA for different polyhedra, as a function of generation number. The dashed lines represent the ABL and ABA values, as reported in Table~\ref{t3}.}
\label{fig9}
\end{figure} 

\section{Convergence and refinement}
\label{conv}
Note that, in order to establish the validity of the method, we set the number of generations to a relatively high value in this work. Although having more structures in the database is expected to improve the quality of statistical analysis, but it increases the computation time considerably. According to our estimate, total number of generations required to obtain the desired result is less than ten, both for single and multiple component system. Initial generations are expected to have some ``undesired'' structures, which are going to be eliminated in due course of evolution. This is shown in Fig~\ref{fig9}(a), where the number of ``undesirable'' cation-cation and anion-anion bonds, present initially, are found to diminish in the subsequent generations. As shown in the figure, initial generations have significant number of oxygen atoms within 1.4 \AA~ ($\sim15$\% higher than O$_2$ molecule bond-length) distance of each other, which are non-existent beyond 10$^{th}$ generation and we observe same trend in case of metal atoms. Accordingly, ABL and ABA values converge approximately by 6$^{th}$ to 10$^{th}$ generations [see Fig~\ref{fig9}(b)]. Note that, structures generated by evolutionary method are similar to as-quenched amorphous structures obtained via melt-quench technique using MD simulations and they can be further refined by annealing at suitable temperature by using \textit{ab initio} MD calculations.       

\section{Conclusion}
\label{concl}
In conclusion, we have successfully generated amorphous structures by using evolutionary algorithm, producing similar results as predicted by popularly used \textit{ab initio} MD based melt-quench technique.  The method is effective for single component (a-Si), as well as multi-component (a-IGZO) systems. In particular, we find that characteristic structural parameters like average bond length and bond angle are within $\sim 2\%$ to those reported by ab initio MD calculation. The strength of EA lies in it's ability to eliminate ``undesirable'' structures, similar to Darwin's theory of evolution based on natural selection. For example, we have observed that in case of \textit{ab initio} MD, if anion-anion bonds are formed during melting, they are inevitably found in the quenched structure, almost appearing like a molecule entrapped in a solid. Such structures, although present initially, do not survive in the later stages of evolutionary search. Moreover, considering the fact that the search converges by approximately 6$^{th}$ to 8$^{th}$ generation, the evolutionary method is not going to be very expensive in terms of computational resources also, than compared to the DFT based MD calculations.
    
\section{Acknowledgement}
SB acknowledges funding from IITK initiation grant. Authors thank CC IITK for providing HPC facility.

\bibliography{references}
\end{document}